# Use of Dynamical Systems Modeling to Hybrid Cloud Database

**Evgeniy V. Pluzhnik, Evgeny V. Nikulchev**
Moscow Technological Institute (WTU), Moscow, Russia
Email: e.pluzhnik@gmail.com, nikulchev@mail.ru

## ABSTRACT

In the article, an experiment is aimed at clarifying the transfer efficiency of the database in the cloud infrastructure. The system was added to the control unit, which has guided the database search in the local part or in the cloud. It is shown that the time data acquisition remains unchanged as a result of modification. Suggestions have been made about the use of the theory of dynamic systems to hybrid cloud database. The present work is aimed at attracting the attention of specialists in the field of cloud database to the apparatus control theory. The experiment presented in this article allows the use of the description of the known methods for solving important practical problems.

## 1. Introduction

Under cloud computing, as a rule, understanding Internet services provided by specialized data centers in the form of hardware and software system, or a distributed computing system consisting of a set of interconnected virtual machines, allows dynamic compute resources to provide a certain level of service [1].

Currently, cloud services are evolving, and there is a question about the transfer efficiency of existing systems and databases in the cloud. Obviously, to get a win, it is necessary to develop criteria for the transfer efficiency systems. The disadvantage of systems that operate in the clouds is the attachment to the communication channels. This paper describes an experiment for proving the opposite of a certain class of systems. The object of investigation used semi-structured data with a large database. In the scientific and educational environment, as a rule, such data—these articles, tutorials, tests, etc., are widely used in systems [2].

The purpose of the article is to adequately describe the processes in cloud databases in the choice of the mathematical apparatus.

Publications in the field of query optimization in cloud data centers are conducted in the last two or three years. It may be noted in these studies [3-6]. In connection with the optimization of queries, there are quite a number of problems: problems of query transformation to a more effective non-procedural representation (logic optimization), the problem of choosing a set of alternative procedural query execution plans, and problems of cost estimates for the selected query execution plan, etc. The problems associated with the logical query optimization have created a direction called semantic optimization. So many researchers study problems on valuations of procedural query execution plans.

To optimize queries, model graph theory, algorithm theory and other methods of discrete mathematics were traditionally used. They cannot get the correct assessment of conformity of theoretical research into practical implementations in the cloud technologies. This is largely due to the fundamentally different building information systems—it is not only the distributed data warehouse, but also the use of virtualization with dynamic reallocation of resources, the use of communication channels with different bandwidths for query processing, various platform features. All these have led to the search for new tool's description of database management systems.

State space representation in the form of a system of equations is not very known in the art to optimize the database and therefore not covered by these experts. The article therefore has a slightly unusual structure. It begins with an example which cannot be described with the help of traditional formal models and optimization tools. And only then it provides a brief overview of the successful use of dynamic models for cloud resource management. Our proposal to use for the design and optimization of the database is made new.

The article has the following structure. The first section presents the results of an experiment, which show the ordinary practical result of cloud technologies. This result cannot be explained by classical theories in the field of search queries to databases. From the point of view of the main provisions of the models based on graph theory and the theory of algorithms, the imposition of the database to the external system must include additional time costs associated with the increase in complexity of the query through the add-on module, as well as the loss of the data transfer. This increase was not observed in this experiment.

The second section of the proposed use is as a mathematical model in the dynamic representation of state space with feedback. Control theory was used in telecommunication systems recently. Our team of dynamic models has been used for solving problems of modeling of network traffic (this model is given as an example in the article [7]).

In conclusion, the optimization problem formulated queries to databases in the cloud, which can be solved by the proposed system.

## 2. Experiment

The experiment represents the production of databases in a hybrid cloud based on a complex search query. The results were compared with the requests of the local client-server system with the same subject databases.

In the table "Articles" are stored articles, article size from 100 KB to 3 MB. In the table "Authers" provides data to the author. Table "Author of Articles" website links with the article. One article may be one main author and several co-authors. In the test load can be from 0 to 9 coauthors. The article shall be in the format docx. Size articles: a maximum—3 MB, Minimum—0.1 MB.

In a local database (articlesLocal) data about authors and articles are stored in a relational database MS SQL Server (see **Figure 1**). The occupied memory on the database server (articlesLocal) 26667,25 MB.

The structure of the hybrid database is shown in **Figure 2**.

In a hybrid DB (articlesHybrid) information about the authors and articles in the local database to MS SQL Server, and the article body—cloud storage Azure Storage. The occupied memory on the database server (articlesHybrid) 47,08 Mb, in the cloud, about 27 GB.

Interaction with the local database is organized by the following way:

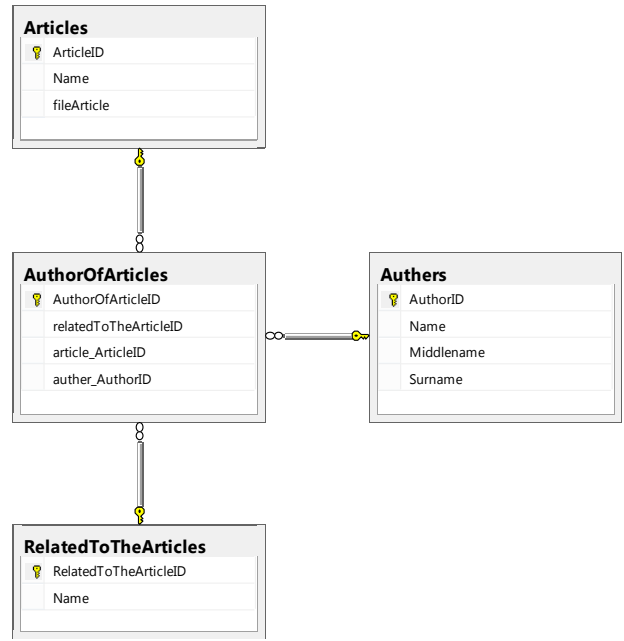

**Figure 1. Database "articlesLocal".**

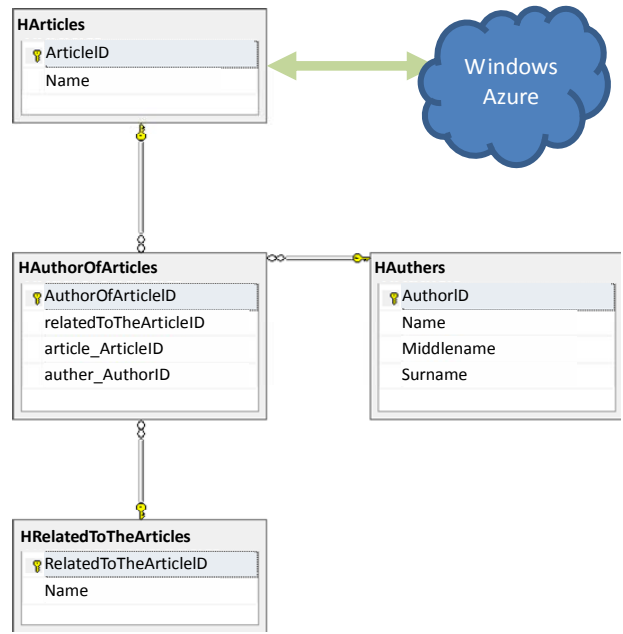

**Figure 2. Database "articlesHybrid".**

1) The client application accesses the service to request articles for the parameter as a parameter was elected by the author of the article.
2) Service calls to the database to SQL Server.
3) SQL returns metadata and text articles service.
4) The service sends the result to the client request.
Interaction with the local database as follows.
Working with hybrid storage:
1) The client application accesses the service request articles of the parameter (a parameter was chosen author

of the article).

2) Service calls to the database to SQL Server.

3) SQL returns service metadata of articles and addresses to which the texts are in Windows Azure.

4) The client receives from the service metadata of articles and addresses.

5) According to the obtained addresses the client accesses the cloud storage and receives text articles.

**Results**

Tests were carried out queries to databases on mining of articles. The results of the first experiment are shown in **Table 1**, **2** and in **Figure 3**, the second in **Table 3**, **4** and **Figure 4**.

The experiment demonstrates the effectiveness of cloud storage for systems using semistructured database.

Based on the results we can conclude that, for the time evaluation of database queries application of graph theory will not be effective. Resources used in the processing of queries on the local server structure other than in an external cloud storage.

## 3. Used to Control System Theory for Cloud Database

### 3.1. Overview

It is proposed to use to evaluate the performance of systems operating in hybrid environments, to use the terminology and methods of system analysis. That certainly is not new, at the levels of service PaaS [1] cloud infrastructure is widely used concept of automatic control, automatic allocation of resources using the methods of dynamics.

**Table 1. Test 1—queries to local databases.**

| Number of records in a query | Number of articles extracted | Records per second | Time of receipt of all articles, sec. |
|---|---|---|---|
| 1 | 100 | 0.908074 | 110,1231453 |
| 2 | 200 | 0.335978 | 297,6383233 |
| 3 | 300 | 0.21054 | 474,969978 |
| 4 | 400 | 0.150936 | 662,533218 |

**Table 2. Test 1—queries to hybrid databases.**

| Number of records in a query | Number of articles extracted | Records per second | Time of receipt of all articles, sec. |
|---|---|---|---|
| 1 | 100 | 0,826204 | 121,0354227 |
| 2 | 200 | 0,329768 | 303,2433013 |
| 3 | 300 | 0,208225 | 480,2488007 |
| 4 | 400 | 0,149446 | 669,136332 |

**Table 3. Test 2—queries to hybrid databases.**

| Number of records in a query | Number of articles extracted | Records per second | Time of receipt of all articles, sec. |
|---|---|---|---|
| 1 | 100 | 0,564751 | 177,0691164 |
| 2 | 200 | 0,257273 | 388,6920493 |
| 3 | 300 | 0,169366 | 590,4366272 |
| 4 | 400 | 0,124121 | 805,6650491 |
| 5 | 500 | 0,094158 | 1062,042418 |
| 6 | 600 | 0,073867 | 1353,779514 |
| 7 | 700 | 0,061519 | 1625,511459 |
| 8 | 800 | 0,052098 | 1919,462269 |
| 9 | 900 | 0,044821 | 2231,100424 |
| 10 | 1000 | 0,038972 | 2565,93466 |

508  E. V. PLUZHNIK, E. V. NIKULCHEV

**Table 4. Test 2—queries to local databases.**

| Number of records in a query | Number of articles extracted | Records per second | Time of receipt of all articles, sec. |
|---|---|---|---|
| 1 | 100 | 0,58732 | 170,2648023 |
| 2 | 200 | 0,26088 | 383,3185936 |
| 3 | 300 | 0,170694 | 585,8428519 |
| 4 | 400 | 0,125036 | 799,76944 |
| 5 | 500 | 0,094675 | 1056,250566 |
| 6 | 600 | 0,074208 | 1347,565385 |
| 7 | 700 | 0,061759 | 1619,206815 |
| 8 | 800 | 0,052273 | 1913,022702 |
| 9 | 900 | 0,044951 | 2224,643836 |
| 10 | 1000 | 0,039075 | 2559,199781 |

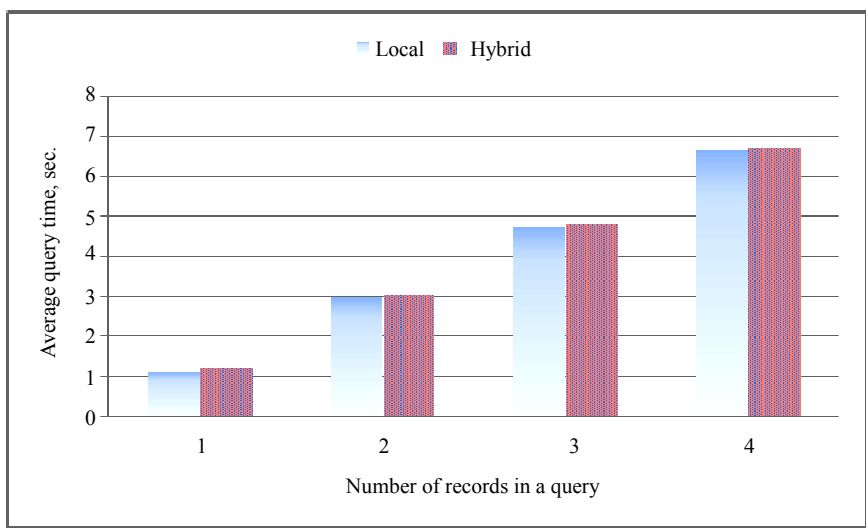

**Figure 3. Results of 1 test.**

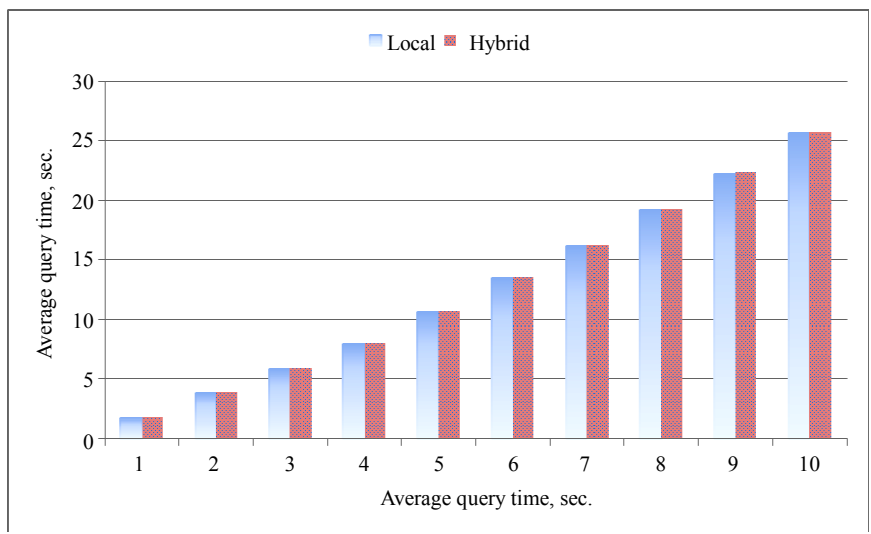

**Figure 4. Results of 2 test.**

For information systems are intuitive terminology of structural complexity, observability, reachability of these information systems. [8,9, etc.]. That means all of the many well-developed tools and control theory can be applied to study the parameters of information systems.

Given the logic of the cloud computing is difficult to determine the effectiveness of the algorithms. Virtualization and scalability of resources may unexpectedly greatly speed up the algorithms, as in our experimental example. That's a probably will help identify methods of system identification.

In general, it can be noted that the transfer system is necessary to develop a stable structure, the study of the characteristics of reliability, observability. Decomposition of the local and cloud components should be based on the methods of structural complexity of systems analysis.

The dynamic descriptions in the form of approximation of differential equations with the vector control were used in the different tasks that are close to the considered in the article. In [10] the processes that include real-time embedded systems. You can also note the earlier article for managing web servers [11], and virtualized data centers [12].

In previous works, such as [13], an approach that assumes that the cloud can be modeled as a Multi-Input-Multi-Output (MIMO), the system is implemented for capacity control in the cloud. However, the design of the regulars and identification systems do not allow the full use of these ideas. In the work [14] proposed to use Single-Input-Single-Output (SISO) systems with the introduction of the notion proportional thresholding.

The original paper [15] is devoted to the construction of a dynamic model of linked servers in the cloud. The important results of this work are the introduction of the concept of positive feedback, the theoretical proof of the stability of the model, the use of passive systems.

The closest in terms of models of our article [16]. This article describes how to identify patterns in the problem space and formulated the principles of the use of automatic control cloud resources.

### 3.2. Main Provisions

Let us consider basic positions which presence constitute grounds for an approach to the control system theory and dynamic models.

*Transients*. Consider a distributed database in the cloud. In cloud databases exist feature is that there are software or hardware routing. They store data on how the virtual machine on which pieces of data is stored, how much is in the public, private cloud which determines the level of security and other service information. In fact, even the same type of incoming query is converted in each case to a different scenario and route maps and depends on the model structure, and communication channels.

Input and output parameters for the study are the parameters characterizing the cloud computing system resources—CPU usage is at a given time, channel load, the control signals on the state of the virtual machines and clusters. In some cases, you can consider the input of a query that can be measured, for example, in the perfect disjunctive normal form.

An important assumption! The system cannot be considered only at a single query, there is a constant stream of requests. The system can be retrieved, then released. Therefore the state of the system at time $t + 1$ depends on the query, and the current state at time $t$. In addition, there are natural constraints imposed by the width of channels, the number of free processors, the amount of free memory, etc.

As you know, in control theory it is transient response of the system to a single signal. A single signal is usually scaled signal outputting system to normal operation, i.e. changes settings when turned on. It is easy to get in a situation where the "bad structure of the clouds" will be oscillatory processes. It will be in a situation where the growth will stimulate queries rotation system increases and decreases in resources. Perhaps monotonous output to normal. All of this means that you must modeling and solution of the classical problems for which was established control theory—stability analysis in the design of feedback systems, the synthesis of optimal controllers, control software.

*Feedback*. The presence of feedback, first defined by the need to consider the current state of the virtual machine workload. As was done in [2]. Secondly for distributed databases final data are delivered to the client via the cloud environment, and information about the end of the query to be delivered to the central system. It is understood that the presence of more feedbacks charged already narrow communication channels and may increase the processing time. This is one more argument in favor of the control systems for the optimization of feedback.

Feature of the cloud is positive feedback.

*Basic model*. The base model offers classic model is approximated by the following difference equations,

$$x(t+1) = Ax(t) + Bu(t),$$

where $x = (x_1, x_2, \cdots, x_n)^T$ – n-dimensional vector of the system states under given constraints $x \in X \subseteq R^n$, $u = (u_1, u_2, \cdots, u_m)^T$ – m-dimensional vector of controls under given constraints $u \in U \subseteq R^m$, $t$ – discrete time instan.

In previous articles on the use of the control CPU, control of power in the data center and cloud computing [14,17], a linear stationary system implementation found to be adequate.

It should be admitted, however, that in some cases, and in the case of non-linear soedli their identification is justified [7,18].

*Structural connections*. It is natural that when you transfer to the cloud structure of the databases and route data retrieval becomes difficult. In the simple example given in the first part of the beginning of this article was introduced by an additional block that separates the search request in the private and public cloud. We do not want to overload the informative presentation so we will not give your example. Here is another example taken from [13], which is shown in **Figure 5**. If we draw the required feedbacks, it will overload picture. So even at the top level representation of cloud databases becomes difficult to follow without mathematical methods for the structural stability of the system. Therefore connected to the structure of the system requires the solution of tasks, known for the theory of management—sustainability assessment framework, assessment of structural complexity, etc.

## 4. Conclusions

In the article an experiment is aimed at clarifying the transfer efficiency of the database in the cloud infrastructure. The system was added to the control unit, which has guided the database search in the local part or in the cloud. It is shown that, as a result of transport, the time data acquisition remains unchanged.

Suggestions have been made about the use of the theory of dynamic systems for the analysis of information systems which use distributed resources and conveyor principle of data processing.

The present work is aimed at attracting the attention of specialists in the field of cloud database to the apparatus control theory. The experiment presented in this article allows the use of the description of the known methods for solving important practical problems. In conclusion, it allows to formulate them.

Investigation of the *stability of systems*—the most important solution for the presence of feedback in distributed systems with disabilities through the communication channel, was limited by safe access to the data, as well as scalable resources under peak load.

The study of *controllability*—for complex structured systems, the problem of controllability in a given time interval, and the task of reachability and observability are not obvious without the pilot study; the use of the proposed model allows the use of famous mathematical tool solutions.

*Designing* effective systems—for large databases with complex levels of access and distributed in different systems with different hardware and software support, as it is observed in a hybrid cloud infrastructure.

*Optimization*—in terms of predetermined criteria of quality construction of effective systems. Modeling can occur both at the design stage, and the step change of the existing systems, when requirements were changed.

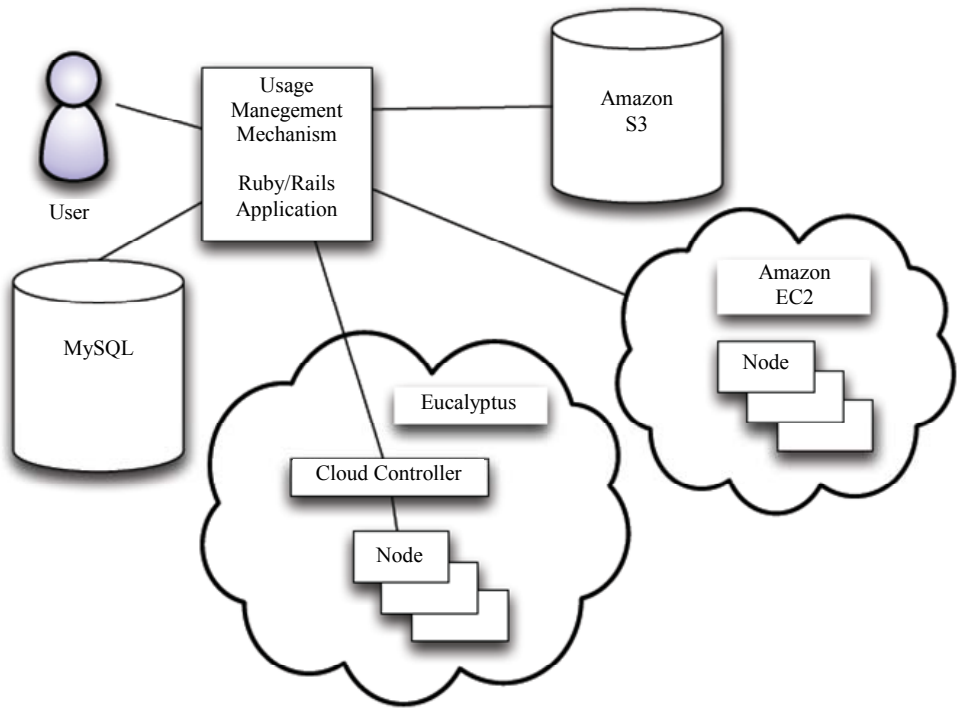

**Figure 5. Example cloud database [13].**

For example, transfer system in the cloud environment is an interesting and urgent task.

## Appendix. Querying the Database

### A1. Hardware

The client computer from which the requests were made:
 Type—Notebook ASUS U36SD,
 Processor—Core i5 2500 MHz, Processor core - 2,
 Memory—4096 MB DDR3 1333 MHz,
 Storage capacity (HDD) 750 GB,
 Drive interface—Serial ATA

The server on which the local database and the local part of the hybrid storage:
 Processor—Intel Atom d525 1800 MHz, Proc. core - 2,
 Memory—4096 MB DDR3,
 Volume drive (HDD)—750 GB,
 Drive interface—Serial ATA.
 Internet connection—Channel capacity 100 Mbit/s
 Cloud storage: Windows Azure,
 Version storage—Locally excessive,
 Territory server locations—Eastern Europe.

### A2. Software

Client—installed operating system is Win 8 64,
 Application winForm, written in C # by using Language-Integrated Query LINQ.
 Server—installed operating system is Windows Server 2012, database server - MS SQL Server 2012.

### A3. Local Database

Request for 10 articles selected author.
 LINQ:
 var auther = bd. Authers. Include ("Author Of Articles. article"). First (item => item. AuthorID == autherID);
var article = auther. Author of Articles. Take (countArticle);
 SQL:
 SELECT TOP **10** * FROM [articlesLocal]. [dbo]. [Authers]
  JOIN [articlesLocal].[dbo].[AuthorOfArticles]
   ON [AuthorOfArticles].[auther_AuthorID] =
      [Authers].[AuthorID]
  LEFT JOIN [articlesLocal].[dbo].[Articles]
   ON [AuthorOfArticles].[article_ArticleID] =
      [Articles].[ArticleID]
      WHERE [Authers].[AuthorID] =
**VARautherID**

### A4. Hybrid Cloud Database

Request for 10 articles selected author.
 LINQ
 var auther = bd. Authers. Include("Author Of Articles. article"). First (item => item. AuthorID == autherID);
 var article = auther. Author of Articles. Take (countArticle);
 SQL
 SELECT TOP **10** * FROM [articlesHybrid] [dbo]. [HAuthers]
  JOIN [articlesHybrid].[dbo].[HAuthorOfArticles]
   ON [HAuthorOfArticles].[auther_AuthorID] =
      [HAuthers].[AuthorID]
  LEFT JOIN [articlesHybrid].[dbo].[HArticles]
   ON [HAuthorOfArticles].[RelatedToTheArticleID] =
      [HArticles].[ArticleID]
      WHERE [HAuthers].[AuthorID] =
**VARautherID**

Request to the cloud storage to extract text articles on the specified address

```
public Stream GetRecord(string id, string TableName = "Article")
    {
      try
       {
  CloudBlobContainer conteiner = Client. GetContainerReference (TableName. ToLower());
  CloudBlockBlob blod = conteiner. Get Block BlobReference (id. To Lower());
            return blod.OpenRead();
       }
       catch (Exception ex)
       {
     MessageBox.Show("error: " + TableName + " " + id.ToString() + " " + ex.Message);
       }
       return null;
    }
    private static CloudBlobClient client;
    public static CloudBlobClient Client
    {
       get
       {
          if (client == null)
          {
CloudStorageAccount storageAccount = CloudStorageAccount.Parse
          (contectionString.ToString());
client = storageAccount.CreateCloudBlobClient();
          }
          return client;
       }
    }
```